\documentclass[fleqn,10pt]{wlscirep}

\usepackage{graphicx}
\usepackage{lineno}
\usepackage{url}

\title{Amino Acid Chiral Selection Via Weak Interactions in Stellar Environments: Implications for the Origin of Life}

\author[1,2,*]{Michael A. Famiano}
\author[2,3]{Richard N. Boyd}
\author[2,4,5+]{Toshitaka Kajino}
\author[4+]{Takashi Onaka}
\author[6+]{Yirong Mo}
\affil[1]{Dept. of Physics and Joint Institute for Nuclear Astrophysics, Western Michigan Univ., 1903 W. Michigan Avenue, Kalamazoo, MI 49008-5252 USA}
\affil[2]{National Astronomical Observatory of Japan, 2-21-1 Mitaka, Tokyo 181-8588 Japan}
\affil[3]{Dept. of Physics, Dept. of Astronomy, The Ohio State Univ., Columbus, OH 43210 USA}
\affil[4]{Dept. of Astronomy, Graduate School of Science; Univ. of Tokyo, 7-3-1 Hongo, Bunkyo-ku, Tokyo 113-0033 Japan}
\affil[5]{School of Physics and Nuclear Energy Engineering, Beihang Univ. (Beijing Univ. of Aeronautics and Astronautics) Beijing 100083, P.R. China}
\affil[6]{Dept. of Chemistry, Western Michigan Univ., 1903 W. Michigan Avenue, Kalamazoo, MI 49008-5252 USA}
\affil[*]{michael.famiano@wmich.edu}

\affil[+]{these authors contributed equally to this work}

\begin{abstract}
	Magnetochiral phenomena may be responsible for the selection of chiral states of biomolecules in
	meteoric environments. 
	For example, the Supernova Amino Acid Processing (SNAAP) Model was proposed previously as a possible mode of magnetochiral 
	selection of amino acids by way of the weak interaction in strong magnetic fields.  
	In earlier work, this model was shown to produce an   
	enantiomeric excess ($ee$) as high as 0.014\% for alanine. 
	In this paper we present the results of molecular quantum chemistry calculations 
	from which $ee$s are determined for the $\alpha$-amino acids 
	plus isovaline and norvaline, which were found to
	have positive $ee$s in meteorites. Calculations are performed for both isolated and aqueous states. In some cases, the 
	aqueous state was found to produce larger $ee$s reaching values as high as a few percent under plausible conditions.  
\end{abstract}

\begin{document} 

\flushbottom

\maketitle
\thispagestyle{empty}

\section*{Introduction}
Recent work on the SNAAP Model of enantiomeric excess ($ee$) production in amino 
acids \cite{boyd10,boyd11,boyd12,famiano14,famiano16,famiano17}
has been shown 
\cite{famiano17} to produce large enough $ee$s to allow it to 
explain how amino acids were created in space and observed in meteorites.  This
	model is initiated via weak interactions in amino acids in strong magnetic and electric fields.
	After an initial enantiomeric excess is created, subsequent autocatalysis and processing could
	be responsible for relative abundances observed in meteorites  \cite{kvenvolden70,bada83,cronin97,cronin98,glavin09,glavin10,herd11}.
We define the enantiomeric excess,
$ee(\%) = \frac{N_L - N_D}{N_L + N_D}\times 100$\%,
where N$_L$ and N$_D$ represent, respectively, the numbers of left- and right-handed amino 
acids in any ensemble. In this paper we present results for determining the $ee$s for 21 amino acids, including proteinogenic $\alpha$-amino acids
(those in which the amine group is bonded directly to the $\alpha$ - or central - carbon atom), as well as isovaline
and norvaline. 
It was shown from meteoritic analyses \cite{herd11} that $ee$s may undergo
	processing
in an aqueous environment.  Thus, we have performed calculations 
for these amino acids both in covalent (gas phase) and 
zwitterionic (aqueous or crystalline) forms, with a few notable examples computed for the amino acids in their cationic or anionic forms.    These include histidine, asparagine, and lysine, which are cationic in solution, as well as aspartic and glutamic acids, which are anionic in solution. In either case, the amino acids could be in an
	aqueous environment.

Following the formation of an amino acid enantiomeric excess
in meteorites, these meteorites may have then seeded the primordial earth.  The encased amino acids would have 
undergone autocatalysis either prior to terrestrial impact
or during subsequent processing on the early earth.    Autocatalytic mechanisms have been postulated to couple a small enantiomeric excess to 
homochirality.  In some models, the pre-existing
chiral bias would have been destroyed in favor of a chiral
bias produced in the primordial environment \cite{gleiser08}.
This and other models may have been enhanced or implemented via
statistical fluctuations in the environment with enough amplitude to drive the environment to a specific 
chirality \cite{hochberg09,jafarpour17}. 

In any case, the origin of amino acid homochirality - whether terrestrial or 
extra-terrestial - has profound implications for the origin of life on earth.  

The details of the SNAAP model, including the mathematical analysis, were presented in earlier 
publications 
\cite{boyd10,boyd11,famiano17}, so only its basics will be presented here. We will then give some 
of the details of the calculations and present the results and conclusions.
\section*{The Supernova Neutrino Amino Acid Processing Model}
In the SNAAP model, described in detail previously 
\cite{boyd12,famiano17}, it is assumed that the amino acids are produced and constrained in meteoroids, 
since this has been found to be the case for those that made their way to Earth \cite{kvenvolden70,bada83,cronin97,cronin98,glavin09,glavin10,herd11}. 
When a supernova explodes, whatever meteoroids are in the vicinity of the nascent neutron star 
will be subjected to the strong magnetic field that results from the core collapse and to an 
intense flux of electron antineutrinos that cool the star. Another possible site \cite{boyd18} is a close binary system comprised of a neutron star and Wolf-Rayet star. An accretion disk would be formed from the material attracted from the WR star to the neutron star. The material in the disk would be processed by the magnetic field from the neutron star, the electric field produced by the motion of meteoroids in the disk through the magnetic field, and the neutrinos from the cooling neutron star.

The electron antineutrinos, hereafter denoted as ``antineutrinos,'' will interact differently 
with the nitrogen nuclei in the amino acids depending on whether the spin of the $^{14}$N (which 
has a spin = 1 in units of Planck's constant divided by 2$\pi$) is aligned or antialigned with 
that of the (spin 1/2) antineutrinos. The relevant reaction is $\bar{\nu}_e +^{14}N\rightarrow ^{14}C+e^+$.
From basic weak interaction nuclear physics it is known \cite{boyd08} that the cross section for 
this reaction if the spins are aligned is about an order of magnitude less than if 
the spins are antialigned, since in the former case, conservation of angular momentum 
requires that the wave function of either the 
$\bar{\nu}_e$ or the $e^+$ provides one unit of angular momentum.
This, along with the orientation of the $^{14}$N spin 
in the magnetic field, produces a selective destruction of  $^{14}$N in one orientation.

Although the $ee$s in any model of amino acid production are smaller than some of those
observed in meteorites, it is generally accepted 
that autocatalysis \cite{soai02,soai14,soai95,meierhenrich08} will drive 
the small amino acid $ee$s to the values observed, and ultimately to Earthly homochirality.  

In the previous work \cite{famiano17}, several effects were investigated that would couple the 
N spin to the alanine molecule. It was shown that the selective destruction of the N would 
indeed produce an $ee$, which was found to be dominated by the 
left-handed chirality observed in meteorites. The maximum $ee$ found was 0.014\%. 
For the present paper we describe calculations 
for the amino acids in covalent, zwitterionic, cationic, and anionic geometries. 
Zwitterionic forms were found to produce 
considerably larger $ee$s in most $\alpha$-amino acids.

It is assumed in the SNAAP model that the relevant objects being processed in the field of 
the supernova explosion would be large enough that some surface material could be ablated away
upon entry into the atmosphere, 
and a meteoroid of a smaller size would remain. However, this object would preserve the $ee$ that 
the original larger object would have had.  Because the interaction cross section for  neutrinos to interact 
with matter is so small \cite{fuller99}, many neutrinos would be able to penetrate to the centers of 
planetary bodies of any size easily.  The extremely large neutrino flux associated with supernova events and neutron stars is what enables a reasonable interaction rate of neutrinos incident on the body.  

The $ee$ produced by such a mechanism would not be altered by ablation heating during atmospheric entry except on the surface.

A critical feature of the SNAAP Model \cite{famiano17} is the 
coupling of the nuclear spin to the molecular chirality. The effects of the molecular 
electronic orbitals and molecular orientation are linked to the nuclear spin via the 
nuclear magnetic shielding tensor, $\mathbf{\sigma}$. Differences in
 the shielding tensor from different molecular chiral states can affect the magnetic 
 fields at the nucleus, and thus the overall spin orientations of the $^{14}$N nuclei with respect 
 to the external field. In this model, external fields can implement this shift via the ``Buckingham
 Effect" \cite{buckingham04,buckingham06,buckingham15}.
Because the mathematics associated with the nuclear-molecular coupling is lengthy, we will 
refer the interested reader to reference \cite{famiano17}. 

\section*{Calculations}
In a previous paper \cite{famiano17}, the values of the nuclear 
magnetic polarizabilities of L- and D-alanine were
calculated following the example of reference \cite{buckingham06}. In the 
current work,
the \texttt{Gaussian} quantum chemistry code \cite{g16} was used to compute the asymmetric
shielding tensor.  Previously, the polarizability was determined via a calculation of the tensor in the presence of external electric fields, which were treated perturbatively.  

Here, we have computed the asymmetric shielding tensors for the $\alpha$-amino acids in both ligand and 
zwitterionic form.  
Electron orbital wave functions were 
optimized in the \texttt{aug-cc-pVDZ} basis \cite{dunning89} using a second-order M{\o}ller-Plesset (MP2)
calculation \cite{mp2} followed by tensor determinations using these orbitals in a DFT 
calculation with the \texttt{B3LYP} hybrid 
functional \cite{b3lyp} and the \texttt{pcS-2}  basis set \cite{jensen08,jensen15}. 

In a coordinate system with axes defined by the fields, the molecular electric dipole moment results in a preferred temperature-dependent orientation
of the molecule with respect to the external electric field and thus a preferred average orientation with respect to the
external magnetic field.  The shielding tensor components 
corresponding to the external fields result in 
chirality-dependent shifts of the magnetic field at the nucleus.  \textit{That is, the molecular orientation is controlled by the external
electric field, the shielding tensor is affected by the molecular orientation in the magnetic field, and the nuclear orientation is affected
by the resulting local magnetic field.}
\subsection*{Effects From the Molecular Electric Dipole Moment.} 
 The influence of the molecular polarization in an
electric field is described below.
Because this model has been described previously \cite{famiano17},  
we describe the main point here, referencing the vector diagram in Figure 6 from Famiano et al. (2018) \cite{famiano17}.

The electric dipole 
moment depends on the molecular shape,
so it is also chiral.  In the presence of 
electric fields, populations of
field-aligned molecules follows a 
Maxwell-Boltzmann distribution \cite{buckingham14,buckingham15};
molecular electronic orbitals have
a preferred orientation in the external electric field.
Likewise, the nuclei have a preferred orientation in the 
external magnetic field. 
The molecular motion in the external magnetic field
then results in a perpendicular electric field. Thus, 
average nuclear orientation can be established relative to the
 molecular orientation. 
Because of the molecular orientation in the external fields, 
a preferred coordinate system is created for the 
shielding tensor; it is not isotropic. 
For an electric dipole 
moment $\mathbf{\mu}_E$, a
nuclear magnetization $\mathbf{M}$, and an
electric field $\mathbf{E}$ , the 
nuclear magnetization shift $\Delta \mathbf{M}_T$ is \cite{buckingham14,buckingham15}: 
\begin{eqnarray}
\label{temp_dep}
\Delta_T\mathbf{M}&=&\frac{1}{6kT}\left[\left(\sigma_{xy}-\sigma_{yx}\right)\mu_{E,z}\right.
\\\nonumber
&~&+\left.\left(\sigma_{yz}-\sigma_{zy}\right)\mu_{E,x}+\left(\sigma_{zx}-\sigma_{xz}\right)\mu_{E,y}\right]
\times\left(\mathbf{M}\times\mathbf{E}\right)
\\\nonumber
~&=&\frac{\mathbf{M}\times\mathbf{E}}{6kT}\eta_M
\end{eqnarray}
	where the ``molecular geometry factor'' $\eta_M$ depends solely on the molecular properties:
	\begin{equation}
	\label{eta_def}
	\eta_M \equiv \left(\sigma_{xy}-\sigma_{yx}\right)\mu_{E,z}+
	\left(\sigma_{yz}-\sigma_{zy}\right)\mu_{E,x}+
	\left(\sigma_{zx}-\sigma_{xz}\right)\mu_{E,y}
	\end{equation}
	Thus the shift in magnetization $\Delta \mathbf{M}$ can be written as a product of a molecular-dependent term, an external environmental-dependent term (temperature and electric field), and the nuclear term $\mathbf{M}$.
	In this model, the magnetization is perpendicular to the electric field, and the relative shift in 
	magnetization is then:
	\begin{equation}
	\frac{\Delta M}{M} = \frac{E}{6kT}\eta_M
	\end{equation}
	The term $\eta_M$ results in a shift in magnetization parallel to the meteoroid velocity in an external magnetic field.
	For $\eta_M>0$, $\Delta M$ is in the direction of the meteoroid velocity, and for $\eta_M<0$, $\Delta M$ is opposite to the meteoroid velocity.  Since $\eta_M$ is chirality-dependent, the resultant magnetization, $\mathbf{M}+\Delta\mathbf{M}$, 
	is chirality dependent, and
	makes an angle $\phi$ with respect to the external magnetic field:
	\begin{eqnarray}
	\tan^{-1}\phi&=& \frac{\Delta M}{M} = \frac{E}{6kT}\eta_M
	\\\nonumber
	\rightarrow\phi&\approx&\frac{E}{6kT}\eta_M
	\end{eqnarray}
A larger value of $\eta_M$ will result in a larger angle, and a greater difference in the component of spins parallel and 
anti-parallel to the spins of the antineutrinos exiting the neutron star.  This will result in a larger $ee$.  The sign of the $ee$ will depend on the value of $\mathbf{M}\times\mathbf{E}_{TS}\cdot\mathbf{v}_{\bar{\nu}}$ and $\eta_M$, where $\mathbf{v}_{\bar{\nu}}$ is the antineutrino
velocity vector, and $\mathbf{E}_{TS}$ is the electric field vector in the meteoroid's rest frame; if both have the same sign, then $ee>$0 (See Figure 6 of Famiano et al. (2018)\cite{famiano17}).  

The above is a result of the thermally averaged electrical polarization of the molecule oriented in a static external electric field. In this case, the anisotropic components of the shielding tensor times the electric dipole moment vector results in a non-zero contribution to the change in magnetization. This non-isotropic effect can exceed the isotropic effects of 
shielding tensor by several orders of magnitude \cite{buckingham06}. 

If the meteoroid is not moving with respect to the magnetic field, the bulk magnetization is $\mathbf{M}$. For a non-zero meteoroid velocity, an 
additional transverse magnetization 
component, $\Delta \mathbf{M}_{L,D}$, is created where the 
subscript indicates the L- or D- chirality. The
chirality-dependent magnetization
creates a shift in the nuclear spin vectors
with populations $N_{\pm,L/D}$. 
The magnetization shift then results in an angle between
the nitrogen spin and the antineutrino spin which is
chirality-dependent.
This means the reaction rates between
 antineutrinos and nitrogen vary 
with chirality, because nitrogen in each chiral state has a different spin alignment.

The reaction rate ratio for antineutrino spins parallel and anti-parallel to the $^{14}$N spin component is estimated \cite{famiano17}:
\begin{equation}
\label{rel_destruction}
\frac{R_p}{R_a} = \frac{1-\cos\Theta_p}{1-\cos\Theta_a}=\frac{1-\sin\phi}{1+\sin\phi}\approx\frac{1-\phi}{1+\phi}
\rightarrow \frac{1-\frac{E}{6kT}\eta_M}{1 + \frac{E}{6kT}\eta_M}.
\end{equation} 
 For the situation described, a positive $\eta_M$
will result in a positive $ee$ as the destruction rates for L-amino acids are smaller. (For 90$^\circ<\phi<$180$^\circ$,
a negative value of $\eta_M$ will result in an excess of L-amino acids.)

With the above destruction rate ratios the time-dependent spin-state populations for each chirality, $N_{\pm,D/L}$, and the enantiomeric excess of a particular state can now be computed. 
\begin{equation}
ee(\%) = \frac{\left(N_{+,L}+N_{-,L}\right)-\left(N_{+,D}+N_{-,D}\right)}{N_{+,L}+N_{-,L}+N_{+,D}+N_{-,D}}
\times 100\%
\end{equation}
\section*{Results and Conclusions}
The value of $\eta_M$ was computed for both the ligand and zwitterionic forms.  
Ligand geometries were taken from the Protein Data Bank \cite{lala}, and zwitterionic geometries were
taken from the Cambridge Structural Database \cite{csd}.

	The relative destruction rate, and hence the resultant $ee$, 
	depends on the geometry factor $\eta_M$.  The computed values of $\eta_M$ are shown in Table \ref{eta_table}
	for both geometries (units in ppm-a.u.). In this table, cations are in bold, and
	anions are underlined. Values separated by commas indicate two geometries in the same
	reference.  Though the ligand geometries generally result in small values of $\eta_M$ of both signs, the zwitterionic geometry generally produces positive $\eta_M$ owing to the larger electric dipole moment of the zwitterionic form. 

In Table \ref{eta_table}, it is seen that nearly all zwitterionic forms have the same sign of $\eta_M$.  Also
	included in this table are the results for the zwitterionic forms after optimizing the literature values as described previously.  For the optimized geometries nearly all values of $\eta_M$ have the same sign, which would result in the same sign of the $ee$ for each amino acid in a single meteoroid
	assuming the same environment throughout.

There are a few notable cases present in this table.  Arginine and histidine both have $\eta_M<0$ in the zwitterionic state.  However, both of these amino acids are likely to have positively-charged side chains in solution.  The values of $\eta_M$ for these cationic states are shown to be positive.   The chosen geometry may have significant uncertainty for this amino acid, however, as it is for the anhydrous monohydrate state, and multiple geometries were listed in the same reference.

Isovaline is another interesting case, as its non-optimized and optimized zwitterionic forms both have $\eta_M<0$.  There are few geometries available for isovaline in the literature, and those available are for
	isovaline monohydrate.   
	However, in the cationic state, which may occur in hydrochloride solutions, $\eta_M$ becomes quite large owing to the significant shift in the dipole moment. In this case, as no literature values for cationic
	isovaline were found, this form was adapted from the monohydrate value and optimized in aqueous solution
	using the same method as the other optimized values in the table.	
	Additional studies of the geometry of isovaline in multiple forms would be welcome and important,
	as this amino acid was found to have a relatively large positive $ee$ in the Murchison meteorite \cite{cronin97}.

With the evaluated values of $\eta_M$, amino acids in Table \ref{eta_table} were simulated in a computer code following
the prescription of reference \cite{famiano17} assuming molecules embedded in meteoroid cores and processed in
high magnetic fields by antineutrinos.  Within the 
meteoroid, an initial population of equal parts D- and L- amino acids is
assumed. The spin-lattice relaxation time for the nitrogen
nucleus was assumed to be
10 ms, which is estimated from previous evaluations \cite{troganis03}. This time could also depend on 
temperature and crystallization state of the
amino acid \cite{nelson03}.  
We assume that the antineutrino and meteoroid velocity vectors are co-linear. 

In this model, one spin-chiral combination is preferentially destroyed  (When the N is converted to C, 
the molecule is no longer an amino acid.), which defines the SNAAP model. 
From Famiano et al. (2018, Figure 6) it is seen that the antineutrino spin alignment with the parallel ($N_+$) nitrogen spin 
for the L- amino acid
 is less than 90$^\circ$ (with respect to the meteoroid velocity vector) while it is greater than 90$^\circ$ 
 in the D- amino acid. The D- amino acid is preferentially destroyed in this configuration as there is a larger spin component opposite the antineutrino spin. For meteoroids moving toward (away from) the central object, 
negative (positive) $ee$s are produced.  

The interaction rate for antineutrino spins
 parallel to the nitrogen spin is parametrized as a factor relative to the inverse relaxation time 
$R_p = f/(2T_1)$. This is the model's only free parameter
and varies 
with the environment in which the model occurs. 
At a distance of 1 AU from a type II SN, 
$f \sim 10^{-9}$ $^($\cite{myra90}$^)$.  For a
neutron star merger $f$ can be $10^{-4}$ 
or much larger depending on meteoroid proximity \cite{perego14}. 
Cooling neutron stars may present another scenario. 
In this case,
$f \sim 10^{-10}$ to $10^{-8}$ $^($\cite{yakovlev95}$^)$. Amino
acids near a neutron star merger are 
expected to be rapidly processed. The
antineutrino reaction rate is expected to be much smaller
for cooling neutron stars. 
Despite this, the flux of antineutrinos could last for
at least $10^5$ years. Processing is possible as long
as the magnetic field is present.  If the meteoroid is
in the vicinity of the neutron star for a few months or 
years, the amino acids will be slowly processed.
Any
situation with an antineutrino flux and high magnetic
fields can result in amino acid processing. 
Equation \ref{rel_destruction} is then 
used to estimate the relative rates of amino acid destruction
for each spin configuration. We also note that the average nuclear spins are nearly perpendicular to the velocity vectors of the 
antineutrinos.

\begin{figure}
	\includegraphics[width=\linewidth]{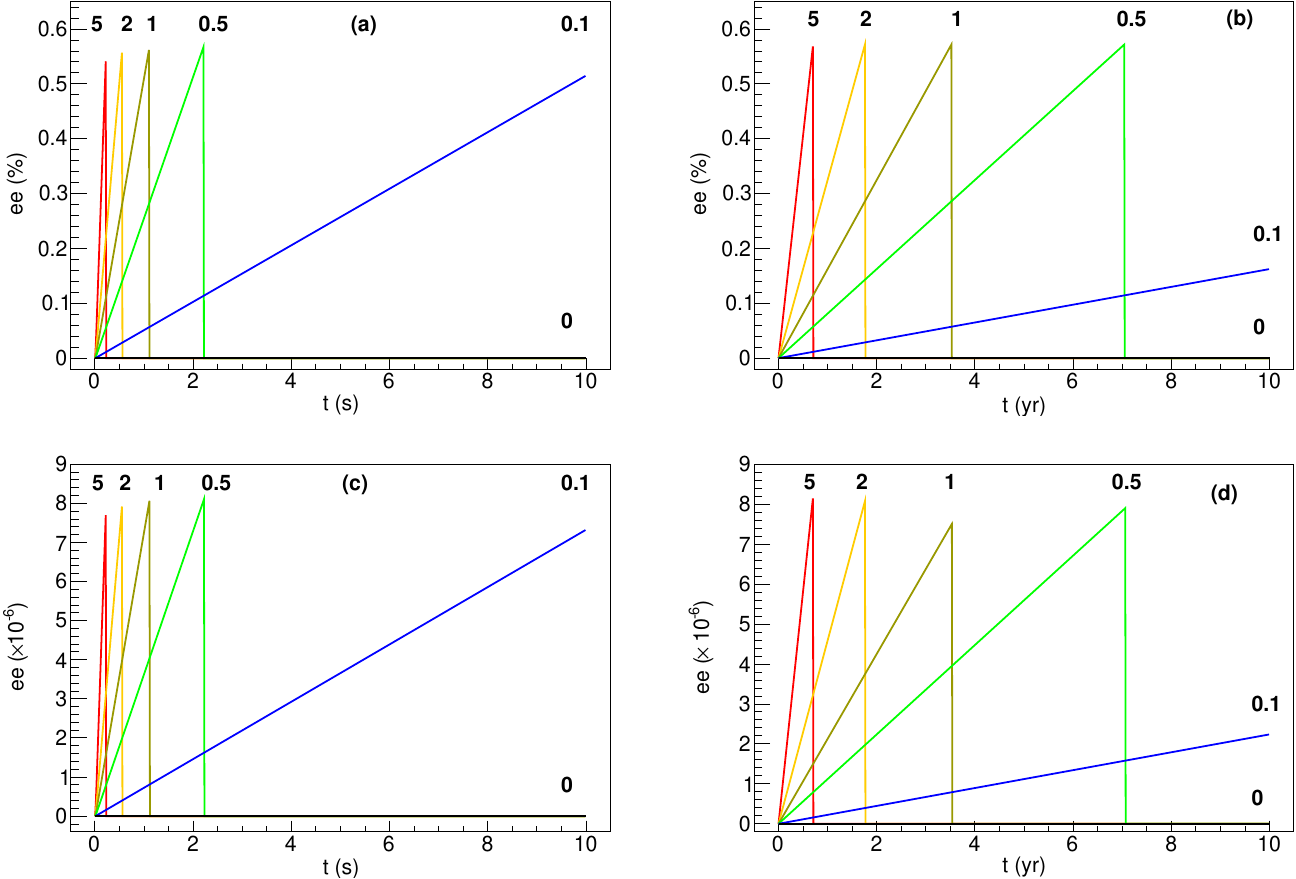}
	\caption{\label{fig7}The isovaline cation $ee$ as a function of time for a meteoroid penetrating the external
		magnetic field for scenarios described in the text, (a) B = 800 T with high f, 
		(b) B = 800 T with low f, (c) B = 30 T with high f, and (c) B = 30 T with low f.
		Values of f are indicated beside the lines. 
		In all cases v = 0.015c. For plots (b) and (d) the values of f should be multiplied by 10$^{-8}$. }
\end{figure}

Figure \ref{fig7} shows the results of
this calculation for cationic isovaline for four scenarios. 
These scenarios show four situations for high and low magnetic fields and high and low antineutrino rates.
The cooling neutron star scenario could be
represented by a magnetic field of 800 T and a meteoroid 
velocity of 0.015c.  For these calculations, the core temperature of the asteroid was assumed to be 10 K. Antineutrino reaction rate 
ratios for this situation are $f=0$, $10^{-9}$, $5\times10^{-9}$, $10^{-8}$, $2\times10^{-8}$, and $5\times10^{-8}$. 
A high-rate, low-field scenario simulates a meteoroid in
proximity to a neutron star merger.  Here $f =$ 0, 0.1, 0.5, 1, 2, and 5;
antineutrino rates that span a large range relative to the
relaxation rate. In this scenario, a magnetic field of 30 T and a velocity of 0.015c is assumed.
The produced $ee$ 
as a function of time is shown in each plot starting from
an $ee$ of zero.
In this model the antineutrino flux controls the 
rate at which the $ee$ changes while the field strengths
control the maximum $ee$ possible.
 Of course, for no external antineutrino interactions ($f=0$), the $ee$ does not change. For all models, the number of
 molecules within a chiral state remains unchanged.

In addition to type II supernova and neutron star mergers, neutron star cooling from a binary star system consisting
of a neutron star and a Wolf-Rayet star \cite{boyd18} may present a possible scenario. With external magnetic fields of 800 T 
at a distance of 500 km and a velocity of 0.015c, the $ee$s can be as high as $\sim$1\%. 
This distance is well within the range of the accretion disk size for a neutron star \cite{boyd18,ludlam17,pringle82}. It is likely that the meteoroid is in an elliptic orbit, so speed would be non-constant.  Additionally, disk viscosity could impact the speed.  However, for a circular orbit 
	at 500 km, the classical orbital velocity would be 0.05c. Thus, we choose a conservative average speed of 0.015c. 

The $ee$ will eventually
drop to zero if the meteoroid spends too much time in the antineutrino flux as all the amino acids will be destroyed. 
A positive $ee$ can be produced for any period of time between 0 and the time at which the $ee$ is a maximum. 
For a cooling neutron star with a very high field and low
antineutrino flux the time to a maximum $ee$ is months to years, the processing time of the meteoroid prior to ejection from the system.  Amino acids will be processed over multiple orbits. A
non-zero $ee$ can be produced during the meteoroid's exposure to the antineutrino flux.

\begin{figure}
	\includegraphics[width=\textwidth]{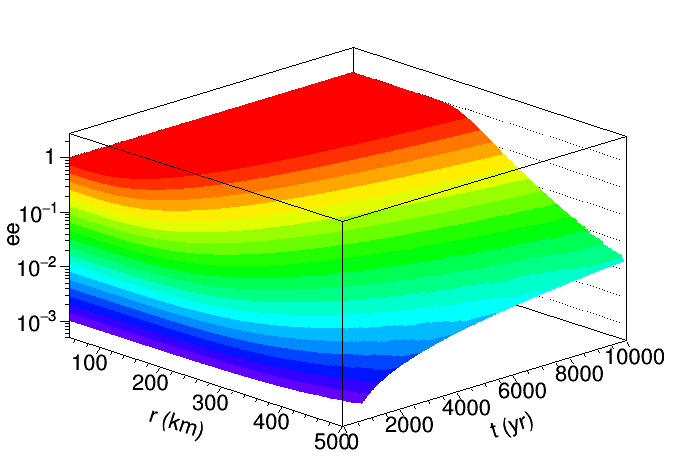}
	\caption{\label{extreme}The $ee$ produced in zwitterionic alanine orbiting a neutron star for various average radii and times. }
\end{figure}

	Because of the wide range in environmental parameters available to this mode, we provide a representation of the 
	$ee$ for zwitterionic alanine as a function of time and
	distance from a neutron star in the NS-WR binary system 
	in Figure \ref{extreme}.  In this 
	system, the meteoroid is presumed to orbit a neutron star
	at the average radius indicated on the plot for the amount of time indicated on the plot.  In this figure, the meteoroid velocity is computed as a function of radius about a neutron star with a surface field of 10$^{15}$ G, a mass of 1 M$_\odot$, and an antineutrino flux of 
	10$^{38}$ s$^{-1}$. Thus, one can use this figure to determine the $ee$ for alanine for an orbit for a certain amount of time.  

The maximum $ee$s computed in this model are approximately proportional to $\eta_M$ from Table 
	\ref{eta_table}.  Thus, we can make a comparison to the relative production of amino acid $ee$s in our 
	model to those found in the Murchison and Murray meteorites \cite{pizzarello00}.  We show the ratios 
	$\eta_M/\eta_{M,ALA}$ in our model compared to the ratios $ee/ee_{ALA}$ found in these meteorites
	\cite{glavin10} in Table
	\ref{ratios}.  
	Data from three sources - CI1 (from the Orgueil meteorite), the CM2 (from the Murchison meteorite), and CR3 (from the Antarctic meteorite QUE 99177) -  samples were examined.
	The CI1 source is known to have a higher degree of post-processed
	aqueous alteration than the CM2 sample, which has a higher level of aqueous alteration than than the CR3 sample.  In the case of 
	glutamic and aspartic acids, the optimized zwitterion geometries were used (though these are generally thought to be anionic in solution). For glutamic acid, the value for the zwitterion is very similar to that of the anion, while for  aspartic acid, the zwitterionic ratio is twice what it would be for the anionic state. 
	Negative values indicate $ee$s with a sign opposite to those of alanine using the molecular geometries in 
	Table \ref{eta_table}.  We see that the presumed ligand and zwitterion computed $ee$ values are not consistent with
	observations.  However, after optimizing these geometries for production in aqueous solutions, the produced $ee$s 
	are closer to those found in the Murchison meteorite \cite{herd11}, with values that appear to more closely
	match those of the CI1 sample, with the exception of glutamic acid, with a lower value.  It seems that an aqueous solution is necessary 
	during or prior to processing.  Certainly the geometry is important in considering this 
	production.  Also, amino acids bound to larger molecules may exhibit shifts in $\eta_M$ as 
	well.  Further, the limited geometries available for isovaline may be relevant as the geometry used here was 
	for crystalline isovaline monohydrate, which we subsequently optimized in aqueous solution.  More study is 
	warranted for this molecule.
	
Here, we point out that - in many cases - 
		the uncertainties in the measurements can be large.  We also note that the range of values between the two samples is quite large. However, the predicted values are similar to the mean $ee$s observed.
		
Any differences may be due to differences between how the amino acids are
		processed in the SNAAP model and any subsequent post-processing through aqueous alteration
		\cite{glavin10}, racemization, heating, and solution effects.  In the current model,
		the amino acids are assumed to be racemic prior to processing.  Processing takes
		a very short amount of time compared to any subsequent racemization and delivery.  Because
		amino acids have different racemization times, subsequent processing of the meteorite
		will alter the initial ratios \cite{cohen00}.  This has been used in the past as a dating mechanism.  Given the variability in racemization times with environment, perhaps it is possible
		to infer meteoric age and subsequent environmental processing in this model.
	
Cumulative environmental effects (e.g., pH, temperature, solvent, etc.) were accounted for in the current study by assuming 
		specific geometries.  Amino acid geometries were taken from the CSD database where
		appropriate \cite{csd}.  For geometry optimization, the geometry references indicated in Table \ref{eta_table} were optimized assuming a neutral 
		water solution with the Polarizable Continuum Model (PCM) \cite{pcm}.  This was independent of any possible subsequent freezing or processing.  Obviously, this is a simplification, and future work will concentrate on more accurate geometry effects.  Thus, care is warranted in comparing values in Table
		\ref{ratios}.
	
	Finally other effects may shift values produced in this model after processing. 
	Subsequent racemization - which occurs at different rates for different amino acids - and 
	auto-catalysis have not been considered in our present model. 
	Other effects may also impact the results. Complex trajectories of the meteoroids may affect the $ee$s, 
	as well as inclusion of the time changing magnetic fields they will encounter. More amino acids may exist 
	in anionic or cationic states, and this may also affect their $ee$s.  Further studies will concentrate on 
	more realistic trajectories as well as the interior environment of the body containing the amino acids.
	None the less, the SNAAP model presents a possibility
	for production of amino acid $ee$s that are approaching the range of those observed in meteorites. 
	
	\section*{Methods}
	The detailed method used to produce the results in this paper are described in
	reference 6 and appendix A therein.  
	
	While the \texttt{DALTON} code\cite{daltonpaper} was used in reference 6, 
	the \texttt{Gaussian}\cite{g16}
	code was used to produce the results in this work using the computational
	method and basis sets described in the text.
	
	\section*{Data Availability}
	The datasets generated during and/or analysed during the current study are available from the corresponding author on reasonable request.
	

\begin{thebibliography}{10}
		\expandafter\ifx\csname url\endcsname\relax
		\def\url#1{\texttt{#1}}\fi
		\expandafter\ifx\csname urlprefix\endcsname\relax\def\urlprefix{URL }\fi
		\providecommand{\bibinfo}[2]{#2}
		\providecommand{\eprint}[2][]{\url{#2}}
		
		\bibitem{boyd10}
		\bibinfo{author}{{Boyd}, R.~N.}, \bibinfo{author}{{Kajino}, T.} \&
		\bibinfo{author}{{Onaka}, T.}
		\newblock \bibinfo{title}{{Supernovae and the chirality of the amino acids}}.
		\newblock \emph{\bibinfo{journal}{Astrobiology}} \textbf{\bibinfo{volume}{10}},
		\bibinfo{pages}{561 -- 568} (\bibinfo{year}{2010}).
		
		\bibitem{boyd11}
		\bibinfo{author}{Boyd, R.~N.}, \bibinfo{author}{Kajino, T.} \&
		\bibinfo{author}{Onaka, T.}
		\newblock \bibinfo{title}{Supernovae, neutrinos and the chirality of amino
			acids}.
		\newblock \emph{\bibinfo{journal}{International Journal of Molecular Sciences}}
		\textbf{\bibinfo{volume}{12}}, \bibinfo{pages}{3432 -- 3444}
		(\bibinfo{year}{2011}).
		
		\bibitem{boyd12}
		\bibinfo{author}{Boyd, R.~N.}
		\newblock \emph{\bibinfo{title}{Stardust, Supernovae and the Molecules of
				Life}} (\bibinfo{publisher}{Springer-Verlag New York}, \bibinfo{year}{2012}).
		
		\bibitem{famiano14}
		\bibinfo{author}{Famiano, M.} \emph{et~al.}
		\newblock \bibinfo{title}{Determining amino acid chirality in the supernova
			neutrino processing model}.
		\newblock \emph{\bibinfo{journal}{Symmetry}} \textbf{\bibinfo{volume}{6}},
		\bibinfo{pages}{909 -- 925} (\bibinfo{year}{2014}).
		
		\bibitem{famiano16}
		\bibinfo{author}{Famiano, M.~A.} \& \bibinfo{author}{Boyd, R.~N.}
		\newblock \emph{\bibinfo{title}{Determining Amino Acid Chirality in the
				Supernova Neutrino Processing Model}}, \bibinfo{pages}{1 -- 17}
		(\bibinfo{publisher}{Springer International Publishing},
		\bibinfo{year}{2016}).
		
		\bibitem{famiano17}
		\bibinfo{author}{Famiano, M.}, \bibinfo{author}{Boyd, R.},
		\bibinfo{author}{Kajino, T.} \& \bibinfo{author}{Onaka, T.}
		\newblock \bibinfo{title}{Selection of amino acid chirality via neutrino
			interactions with $^{14}$n in crossed electric and magnetic fields}.
		\newblock \emph{\bibinfo{journal}{Astrobiology}} \textbf{\bibinfo{volume}{18}},
		\bibinfo{pages}{290 -- 206} (\bibinfo{year}{2018}).
		
		\bibitem{kvenvolden70}
		\bibinfo{author}{{Kvenvolden}, K.} \emph{et~al.}
		\newblock \bibinfo{title}{{Evidence for extraterrestrial amino-acids and
				hydrocarbons in the Murchison meteorite}}.
		\newblock \emph{\bibinfo{journal}{Nature}} \textbf{\bibinfo{volume}{228}},
		\bibinfo{pages}{923 -- 926} (\bibinfo{year}{1970}).
		
		\bibitem{bada83}
		\bibinfo{author}{{Bada}, J.~L.} \emph{et~al.}
		\newblock \bibinfo{title}{{On the reported optical activity of amino acids in
				the Murchison meteorite}}.
		\newblock \emph{\bibinfo{journal}{Nature}} \textbf{\bibinfo{volume}{301}},
		\bibinfo{pages}{494 -- 496} (\bibinfo{year}{1983}).
		
		\bibitem{cronin97}
		\bibinfo{author}{{Cronin}, J.~R.} \& \bibinfo{author}{{Pizzarello}, S.}
		\newblock \bibinfo{title}{{Enantiomeric excesses in meteoritic amino acids}}.
		\newblock \emph{\bibinfo{journal}{Science}} \textbf{\bibinfo{volume}{275}},
		\bibinfo{pages}{951 -- 955} (\bibinfo{year}{1997}).
		
		\bibitem{cronin98}
		\bibinfo{author}{Cronin, J.}, \bibinfo{author}{Pizzarello, S.} \&
		\bibinfo{author}{Cruikshank, D.}
		\newblock \bibinfo{title}{Organic matter in carbonaceous chondrites, planetary satellites, asteroids and comets in \emph{Meteorites and the Early Solar System} (eds. Kerridge, J.F. \& Matthews, M.S.)}
		\bibinfo{pages}{819 -- 857} (\bibinfo{publisher}{University of Arizona
			Press}, \bibinfo{address}{Tucson}, \bibinfo{year}{1998}).
		
		\bibitem{glavin09}
		\bibinfo{author}{{Glavin}, D.~P.} \& \bibinfo{author}{{Dworkin}, J.~P.}
		\newblock \bibinfo{title}{{Enrichment of the amino acid L-isovaline by aqueous
				alteration on CI and CM meteorite parent bodies}}.
		\newblock \emph{\bibinfo{journal}{Proceedings of the National Academy of
				Science}} \textbf{\bibinfo{volume}{106}}, \bibinfo{pages}{5487 -- 5492}
		(\bibinfo{year}{2009}).
		
		\bibitem{glavin10}
		\bibinfo{author}{{Glavin}, D.~P.}, \bibinfo{author}{{Callahan}, M.~P.},
		\bibinfo{author}{{Dworkin}, J.~P.} \& \bibinfo{author}{{Elsila}, J.~E.}
		\newblock \bibinfo{title}{{The effects of parent body processes on amino acids
				in carbonaceous chondrites}}.
		\newblock \emph{\bibinfo{journal}{Meteoritics and Planetary Science}}
		\textbf{\bibinfo{volume}{45}}, \bibinfo{pages}{1948 -- 1972}
		(\bibinfo{year}{2010}).
		
		\bibitem{herd11}
		\bibinfo{author}{{Herd}, C.~D.~K.} \emph{et~al.}
		\newblock \bibinfo{title}{{Origin and evolution of prebiotic organic matter as
				inferred from the Tagish Lake meteorite}}.
		\newblock \emph{\bibinfo{journal}{Science}} \textbf{\bibinfo{volume}{332}},
		\bibinfo{pages}{1304 -- 1307} (\bibinfo{year}{2011}).
		
		\bibitem{gleiser08}
		\bibinfo{author}{Gleiser, M.}, \bibinfo{author}{Thorarinson, J.} \&
		\bibinfo{author}{Walker, S.~I.}
		\newblock \bibinfo{title}{Punctuated chirality}.
		\newblock \emph{\bibinfo{journal}{Origins of Life and Evolution of Biospheres}}
		\textbf{\bibinfo{volume}{38}}, \bibinfo{pages}{499--508}
		(\bibinfo{year}{2008}).
		
		\bibitem{hochberg09}
		\bibinfo{author}{Hochberg, D.}
		\newblock \bibinfo{title}{Mirror symmetry breaking and restoration: The role of
			noise and chiral bias}.
		\newblock \emph{\bibinfo{journal}{Phys. Rev. Lett.}}
		\textbf{\bibinfo{volume}{102}}, \bibinfo{pages}{248101}
		(\bibinfo{year}{2009}).
		
		\bibitem{jafarpour17}
		\bibinfo{author}{Jafarpour, F.}, \bibinfo{author}{Biancalani, T.} \&
		\bibinfo{author}{Goldenfeld, N.}
		\newblock \bibinfo{title}{Noise-induced symmetry breaking far from equilibrium
			and the emergence of biological homochirality}.
		\newblock \emph{\bibinfo{journal}{Phys. Rev. E}} \textbf{\bibinfo{volume}{95}},
		\bibinfo{pages}{032407} (\bibinfo{year}{2017}).
		
		\bibitem{boyd18}
		\bibinfo{author}{Boyd, R.~N.}, \bibinfo{author}{Famiano, M.~A.},
		\bibinfo{author}{Onaka, T.} \& \bibinfo{author}{Kajino, T.}
		\newblock \bibinfo{title}{Sites that can produce left-handed amino acids in the
			supernova neutrino amino acid processing model}.
		\newblock \emph{\bibinfo{journal}{The Astrophysical Journal}}
		\textbf{\bibinfo{volume}{856}}, \bibinfo{pages}{26 -- 30}
		(\bibinfo{year}{2018}).
		
		\bibitem{boyd08}
		\bibinfo{author}{Boyd, R.~N.}
		\newblock \emph{\bibinfo{title}{An Introduction to Nuclear Astrophysics}}
		(\bibinfo{publisher}{The University of Chicago Press, Chicago},
		\bibinfo{year}{2008}).
		
		\bibitem{soai02}
		\bibinfo{author}{Soai, K.} \& \bibinfo{author}{Sato, I.}
		\newblock \bibinfo{title}{Asymmetric autocatalysis and its application to
			chiral discrimination}.
		\newblock \emph{\bibinfo{journal}{Chirality}} \textbf{\bibinfo{volume}{14}},
		\bibinfo{pages}{548 -- 554} (\bibinfo{year}{2002}).
		
		\bibitem{soai14}
		\bibinfo{author}{Soai, K.}, \bibinfo{author}{Kawasaki, T.} \&
		\bibinfo{author}{Matsumoto, A.}
		\newblock \bibinfo{title}{The origins of homochirality examined by using
			asymmetric autocatalysis}.
		\newblock \emph{\bibinfo{journal}{The Chemical Record}}
		\textbf{\bibinfo{volume}{14}}, \bibinfo{pages}{70 -- 83}
		(\bibinfo{year}{2014}).
		
		\bibitem{soai95}
		\bibinfo{author}{Soai, K.}, \bibinfo{author}{Shibata, T.},
		\bibinfo{author}{Morioka, H.} \& \bibinfo{author}{Choji, K.}
		\newblock \bibinfo{title}{Asymmetric autocatalysis and amplification of
			enantiomeric excess of a chiral molecule}.
		\newblock \emph{\bibinfo{journal}{Nature}} \textbf{\bibinfo{volume}{378}},
		\bibinfo{pages}{767 -- 768} (\bibinfo{year}{1995}).
		
		\bibitem{meierhenrich08}
		\bibinfo{author}{Meierhenrich, U.}
		\newblock \emph{\bibinfo{title}{Amino Acids and the Asymmetry of Life: Caught
				in the Act of Formation}} (\bibinfo{publisher}{Springer-Verlag Verlin
			Heidelberg}, \bibinfo{year}{2008}).
		
		\bibitem{fuller99}
		\bibinfo{author}{{Fuller}, G.~M.}, \bibinfo{author}{{Haxton}, W.~C.} \&
		\bibinfo{author}{{McLaughlin}, G.~C.}
		\newblock \bibinfo{title}{{Prospects for detecting supernova neutrino flavor
				oscillations}}.
		\newblock \emph{\bibinfo{journal}{Physical Review D}}
		\textbf{\bibinfo{volume}{59}}, \bibinfo{pages}{085005--1 -- 85005--15}
		(\bibinfo{year}{1999}).
		
		\bibitem{buckingham04}
		\bibinfo{author}{Buckingham, A.}
		\newblock \bibinfo{title}{Chirality in nmr spectroscopy}.
		\newblock \emph{\bibinfo{journal}{Chemical Physics Letters}}
		\textbf{\bibinfo{volume}{398}}, \bibinfo{pages}{1 -- 5}
		(\bibinfo{year}{2004}).
		
		\bibitem{buckingham06}
		\bibinfo{author}{Buckingham, A.} \& \bibinfo{author}{Fischer, P.}
		\newblock \bibinfo{title}{Direct chiral discrimination in nmr spectroscopy}.
		\newblock \emph{\bibinfo{journal}{Chemical Physics}}
		\textbf{\bibinfo{volume}{324}}, \bibinfo{pages}{111 -- 116}
		(\bibinfo{year}{2006}).
		
		\bibitem{buckingham15}
		\bibinfo{author}{Buckingham, A.~D.}, \bibinfo{author}{Lazzeretti, P.} \&
		\bibinfo{author}{Pelloni, S.}
		\newblock \bibinfo{title}{Chiral discrimination in nmr spectroscopy:
			computation of the relevant molecular pseudoscalars}.
		\newblock \emph{\bibinfo{journal}{Molecular Physics}}
		\textbf{\bibinfo{volume}{113}}, \bibinfo{pages}{1780 -- 1785}
		(\bibinfo{year}{2015}).
		
		\bibitem{g16}
		\bibinfo{author}{Frisch, M.~J.} \emph{et~al.}
		\newblock \bibinfo{title}{Gaussian 16 {R}evision {A}.03}
		(\bibinfo{year}{2016}).
		\newblock \bibinfo{note}{Gaussian Inc. Wallingford CT}.
		
		\bibitem{dunning89}
		\bibinfo{author}{Jr., T. H.~D.}
		\newblock \bibinfo{title}{Gaussian basis sets for use in correlated molecular
			calculations. i. the atoms boron through neon and hydrogen}.
		\newblock \emph{\bibinfo{journal}{The Journal of Chemical Physics}}
		\textbf{\bibinfo{volume}{90}}, \bibinfo{pages}{1007--1023}
		(\bibinfo{year}{1989}).
		
		\bibitem{mp2}
		\bibinfo{author}{Jensen, H.}, \bibinfo{author}{Aa, J.},
		\bibinfo{author}{J\o{}rgensen, P.}, \bibinfo{author}{\AA{}gren, H.} \&
		\bibinfo{author}{Olsen, J.}
		\newblock \bibinfo{title}{Second -- order {M}\o{}ller-{P}lesset perturbation
			theory as a configuration and orbital generator in multiconfiguration
			self{-}consistent field calculations}.
		\newblock \emph{\bibinfo{journal}{The Journal of Chemical Physics}}
		\textbf{\bibinfo{volume}{88}}, \bibinfo{pages}{3834 -- 3839}
		(\bibinfo{year}{1988}).
		
		\bibitem{b3lyp}
		\bibinfo{author}{Becke, A.~D.}
		\newblock \bibinfo{title}{Density -- functional thermochemistry. iii. the role
			of exact exchange}.
		\newblock \emph{\bibinfo{journal}{The Journal of Chemical Physics}}
		\textbf{\bibinfo{volume}{98}}, \bibinfo{pages}{5648 -- 5652}
		(\bibinfo{year}{1993}).
		
		\bibitem{jensen08}
		\bibinfo{author}{Jensen, F.}
		\newblock \bibinfo{title}{Basis set convergence of nuclear magnetic shielding
			constants calculated by density functional methods}.
		\newblock \emph{\bibinfo{journal}{Journal of Chemical Theory and Computation}}
		\textbf{\bibinfo{volume}{4}}, \bibinfo{pages}{719 -- 727}
		(\bibinfo{year}{2008}).
		
		\bibitem{jensen15}
		\bibinfo{author}{Jensen, F.}
		\newblock \bibinfo{title}{Segmented contracted basis sets optimized for nuclear
			magnetic shielding}.
		\newblock \emph{\bibinfo{journal}{Journal of Chemical Theory and Computation}}
		\textbf{\bibinfo{volume}{11}}, \bibinfo{pages}{132 -- 138}
		(\bibinfo{year}{2015}).
		
		\bibitem{buckingham14}
		\bibinfo{author}{Buckingham, A.~D.}
		\newblock \bibinfo{title}{Communication: Permanent dipoles contribute to
			electric polarization in chiral nmr spectra}.
		\newblock \emph{\bibinfo{journal}{The Journal of Chemical Physics}}
		\textbf{\bibinfo{volume}{140}}, \bibinfo{pages}{011103--1 -- 11103--3}
		(\bibinfo{year}{2014}).
		
		\bibitem{lala}
		\bibinfo{author}{S, K.} \emph{et~al.}
		\newblock \bibinfo{title}{Pubchem substance and compound databases}.
		\newblock \emph{\bibinfo{journal}{Nucleic Acids Res.}}
		\textbf{\bibinfo{volume}{44}}, \bibinfo{pages}{D1202--13}
		(\bibinfo{year}{2015}).
		
		\bibitem{csd}
		\bibinfo{author}{Groom, C.~R.}, \bibinfo{author}{Bruno, I.~J.},
		\bibinfo{author}{Lightfoot, M.~P.} \& \bibinfo{author}{Ward, S.~C.}
		\newblock \bibinfo{title}{{The Cambridge Structural Database}}.
		\newblock \emph{\bibinfo{journal}{Acta Crystallographica Section B}}
		\textbf{\bibinfo{volume}{72}}, \bibinfo{pages}{171--179}
		(\bibinfo{year}{2016}).
		
		\bibitem{troganis03}
		\bibinfo{author}{Troganis, A.~N.}, \bibinfo{author}{Tsanaktsidis, C.} \&
		\bibinfo{author}{Gerothanassis, I.~P.}
		\newblock \bibinfo{title}{$^{14}${N} {NMR} relaxation times of several protein
			amino acids in aqueous solution—comparison with $^{17}${O} {NMR} data and
			estimation of the relative hydration numbers in the cationic and zwitterionic
			forms}.
		\newblock \emph{\bibinfo{journal}{Journal of Magnetic Resonance}}
		\textbf{\bibinfo{volume}{164}}, \bibinfo{pages}{294 -- 303}
		(\bibinfo{year}{2003}).
		
		\bibitem{nelson03}
		\bibinfo{author}{Nelson, J.}
		\newblock \emph{\bibinfo{title}{Nuclear Magnetic Resonance Spectroscopy}}
		(\bibinfo{publisher}{Pearson Education, New Jersey}, \bibinfo{year}{2003}).
		
		\bibitem{myra90}
		\bibinfo{author}{{Myra}, E.~S.} \& \bibinfo{author}{{Burrows}, A.}
		\newblock \bibinfo{title}{{Neutrinos from type II supernovae - The first 100
				milliseconds}}.
		\newblock \emph{\bibinfo{journal}{Astrophysical Journal}}
		\textbf{\bibinfo{volume}{364}}, \bibinfo{pages}{222--231}
		(\bibinfo{year}{1990}).
		
		\bibitem{perego14}
		\bibinfo{author}{{Perego}, A.} \emph{et~al.}
		\newblock \bibinfo{title}{{Neutrino-driven winds from neutron star merger
				remnants}}.
		\newblock \emph{\bibinfo{journal}{Monthly Notices of the Royal Astronomical
				Society}} \textbf{\bibinfo{volume}{443}}, \bibinfo{pages}{3134--3156}
		(\bibinfo{year}{2014}).
		
		\bibitem{yakovlev95}
		\bibinfo{author}{{Yakovlev}, D.~G.} \& \bibinfo{author}{{Levenfish}, K.~P.}
		\newblock \bibinfo{title}{{Modified URCA process in neutron star cores.}}
		\newblock \emph{\bibinfo{journal}{Astronomy \& Astrophysics}}
		\textbf{\bibinfo{volume}{297}}, \bibinfo{pages}{717} (\bibinfo{year}{1995}).
		
		\bibitem{ludlam17}
		\bibinfo{author}{{Ludlam}, R.~M.} \emph{et~al.}
		\newblock \bibinfo{title}{{A hard look at the neutron stars and accretion disks
				in 4U 1636-53, GX 17+2, and 4U 1705-44 with NuStar}}.
		\newblock \emph{\bibinfo{journal}{The Astrophysical Journal}}
		\textbf{\bibinfo{volume}{836}}, \bibinfo{pages}{140} (\bibinfo{year}{2017}).
		
		\bibitem{pringle82}
		\bibinfo{author}{{Pringle}, J.~E.}
		\newblock \bibinfo{title}{{Accretion discs around neutron stars}}.
		\newblock In \bibinfo{editor}{{Brinkmann}, W.} \& \bibinfo{editor}{{Truemper},
			J.} (eds.) \emph{\bibinfo{booktitle}{Accreting Neutron Stars}}
		(\bibinfo{year}{1982}).
		
		\bibitem{pizzarello00}
		\bibinfo{author}{{Pizzarello}, S.} \& \bibinfo{author}{{Cronin}, J.~R.}
		\newblock \bibinfo{title}{{Non-racemic amino acids in the Murray and Murchison
				meteorites}}.
		\newblock \emph{\bibinfo{journal}{Geoch. et Cosmoch. Acta}}
		\textbf{\bibinfo{volume}{64}}, \bibinfo{pages}{329--338}
		(\bibinfo{year}{2000}).
		
		\bibitem{cohen00}
		\bibinfo{author}{Cohen, B.~A.} \& \bibinfo{author}{Chyba, C.~F.}
		\newblock \bibinfo{title}{Racemization of meteoritic amino acids}.
		\newblock \emph{\bibinfo{journal}{Icarus}} \textbf{\bibinfo{volume}{145}},
		\bibinfo{pages}{272 -- 281} (\bibinfo{year}{2000}).
		
		\bibitem{pcm}
		\bibinfo{author}{Mennucci, B.}
		\newblock \bibinfo{title}{Polarizable continuum model}.
		\newblock \emph{\bibinfo{journal}{Wiley Interdisciplinary Reviews:
				Computational Molecular Science}} \textbf{\bibinfo{volume}{2}},
		\bibinfo{pages}{386--404}.
		
		\bibitem{daltonpaper}
		\bibinfo{author}{Aidas, K.} \emph{et~al.}
		\newblock \bibinfo{title}{{The Dalton quantum chemistry program system}}.
		\newblock \emph{\bibinfo{journal}{WIREs Comput. Mol. Sci.}}
		\textbf{\bibinfo{volume}{4}}, \bibinfo{pages}{269 -- 284}
		(\bibinfo{year}{2014}).
		
		\bibitem{LALNIN59}
		\bibinfo{author}{Woi{\'n}ska, M.}, \bibinfo{author}{Grabowsky, S.},
		\bibinfo{author}{Dominiak, P.~M.}, \bibinfo{author}{Wo{\'z}niak, K.} \&
		\bibinfo{author}{Jayatilaka, D.}
		\newblock \bibinfo{title}{Hydrogen atoms can be located accurately and
			precisely by x-ray crystallography}.
		\newblock \emph{\bibinfo{journal}{Science Advances}}
		\textbf{\bibinfo{volume}{2}}, \bibinfo{pages}{e1600192} (\bibinfo{year}{2016}).
		
		\bibitem{XIYSAA}
		\bibinfo{author}{Yamada, K.}, \bibinfo{author}{Sato, A.},
		\bibinfo{author}{Shimizu, T.}, \bibinfo{author}{Yamazaki, T.} \&
		\bibinfo{author}{Yokoyama, S.}
		\newblock \bibinfo{title}{{{\sc l}-Alanine hydrochloride monohydrate}}.
		\newblock \emph{\bibinfo{journal}{Acta Crystallographica Section E}}
		\textbf{\bibinfo{volume}{64}}, \bibinfo{pages}{o806} (\bibinfo{year}{2008}).
		
		\bibitem{TAQBIY}
		\bibinfo{author}{Courvoisier, E.}, \bibinfo{author}{Williams, P.~A.},
		\bibinfo{author}{Lim, G.~K.}, \bibinfo{author}{Hughes, C.~E.} \&
		\bibinfo{author}{Harris, K. D.~M.}
		\newblock \bibinfo{title}{The crystal structure of l-arginine}.
		\newblock \emph{\bibinfo{journal}{Chem. Commun.}}
		\textbf{\bibinfo{volume}{48}}, \bibinfo{pages}{2761--2763}
		(\bibinfo{year}{2012}).
		
		\bibitem{ARGHCL11}
		\bibinfo{author}{Xian, L.} \& \bibinfo{author}{Gong-xuan, L.}
		\newblock \bibinfo{title}{Conformational changes of arginine dichloride
			dihydrate}.
		\newblock \emph{\bibinfo{journal}{Haxue Yanjiu Yu Yingyong}}
		\textbf{\bibinfo{volume}{19}}, \bibinfo{pages}{474 -- 478}
		(\bibinfo{year}{2007}).
		
		\bibitem{VIKKEG}
		\bibinfo{author}{Yamada, K.}, \bibinfo{author}{Hashizume, D.},
		\bibinfo{author}{Shimizu, T.} \& \bibinfo{author}{Yokoyama, S.}
		\newblock \bibinfo{title}{{{\sc l}-Asparagine}}.
		\newblock \emph{\bibinfo{journal}{Acta Crystallographica Section E}}
		\textbf{\bibinfo{volume}{63}}, \bibinfo{pages}{o3802--o3803}
		(\bibinfo{year}{2007}).
		
		\bibitem{ASPARM09}
		\bibinfo{author}{Flaig, R.}, \bibinfo{author}{Koritsanszky, T.},
		\bibinfo{author}{Dittrich, B.}, \bibinfo{author}{Wagner, A.} \&
		\bibinfo{author}{Luger, P.}
		\newblock \bibinfo{title}{Intra- and intermolecular topological properties of
			amino acids: A comparative study of experimental and theoretical results}.
		\newblock \emph{\bibinfo{journal}{Journal of the American Chemical Society}}
		\textbf{\bibinfo{volume}{124}}, \bibinfo{pages}{3407--3417}
		(\bibinfo{year}{2002}).
		
		\bibitem{LASPRT04}
		\bibinfo{author}{Bendeif, E.-e.} \& \bibinfo{author}{Jelsch, C.}
		\newblock \bibinfo{title}{{The experimental library multipolar atom model
				refinement of {\sc l}-aspartic acid}}.
		\newblock \emph{\bibinfo{journal}{Acta Crystallographica Section C}}
		\textbf{\bibinfo{volume}{63}}, \bibinfo{pages}{o361} (\bibinfo{year}{2007}).
		
		\bibitem{NAGLYB10}
		\bibinfo{author}{Salunke, D.~M.} \& \bibinfo{author}{Vijayan, M.}
		\newblock \bibinfo{title}{{{\sc l}-Arginine {\sc l}-aspartate}}.
		\newblock \emph{\bibinfo{journal}{Acta Crystallographica Section B}}
		\textbf{\bibinfo{volume}{38}}, \bibinfo{pages}{1328--1330}
		(\bibinfo{year}{1982}).
		
		\bibitem{LCYSTN36}
		\bibinfo{author}{Kolesov, B.~A.}, \bibinfo{author}{Minkov, V.~S.},
		\bibinfo{author}{Boldyreva, E.~V.} \& \bibinfo{author}{Drebushchak, T.~N.}
		\newblock \bibinfo{title}{Phase transitions in the crystals of l- and
			dl-cysteine on cooling: Intermolecular hydrogen bonds distortions and the
			side-chain motions of thiol-groups. 1. l-cysteine}.
		\newblock \emph{\bibinfo{journal}{The Journal of Physical Chemistry B}}
		\textbf{\bibinfo{volume}{112}}, \bibinfo{pages}{12827--12839}
		(\bibinfo{year}{2008}).
		
		\bibitem{LGLUAC03}
		\bibinfo{author}{Lehmann, M.~S.} \& \bibinfo{author}{Nunes, A.~C.}
		\newblock \bibinfo{title}{{A short hydrogen bond between near identical
				carboxyl groups in the {$\alpha$}-modification of {\sc l}-glutamic acid}}.
		\newblock \emph{\bibinfo{journal}{Acta Crystallographica Section B}}
		\textbf{\bibinfo{volume}{36}}, \bibinfo{pages}{1621--1625}
		(\bibinfo{year}{1980}).
		
		\bibitem{CAGLCL10}
		\bibinfo{author}{Einspahr, H.}, \bibinfo{author}{Gartland, G.~L.} \&
		\bibinfo{author}{Bugg, C.~E.}
		\newblock \bibinfo{title}{{Calcium binding to {$\alpha$}-amino acids: the
				crystal structure of calcium {\it L}-glutamate chloride monohydrate}}.
		\newblock \emph{\bibinfo{journal}{Acta Crystallographica Section B}}
		\textbf{\bibinfo{volume}{33}}, \bibinfo{pages}{3385--3390}
		(\bibinfo{year}{1977}).
		
		\bibitem{GLUTAM02}
		\bibinfo{author}{Wagner, A.} \& \bibinfo{author}{Luger, P.}
		\newblock \bibinfo{title}{Charge density and topological analysis of
			l-glutamine}.
		\newblock \emph{\bibinfo{journal}{Journal of Molecular Structure}}
		\textbf{\bibinfo{volume}{595}}, \bibinfo{pages}{39 -- 46}
		(\bibinfo{year}{2001}).
		
		\bibitem{LHISTD13}
		\bibinfo{author}{Lehmann, M.}, \bibinfo{author}{Koetzle, T.} \&
		\bibinfo{author}{Hamilton, W.}
		\newblock \emph{\bibinfo{journal}{International Journal of Peptide and Protein
				Research}} \textbf{\bibinfo{volume}{4}}, \bibinfo{pages}{229}
		(\bibinfo{year}{1972}).
		
		\bibitem{HISTCM12}
		\bibinfo{author}{Fuess, H.}, \bibinfo{author}{Hohlwein, D.} \&
		\bibinfo{author}{Mason, S.~A.}
		\newblock \bibinfo{title}{{Neutron diffraction study of {\sc l}-histidine
				hydrochloride monohydrate}}.
		\newblock \emph{\bibinfo{journal}{Acta Crystallographica Section B}}
		\textbf{\bibinfo{volume}{33}}, \bibinfo{pages}{654--659}
		(\bibinfo{year}{1977}).
		
		\bibitem{LISLEU02}
		\bibinfo{author}{G{\"{o}}rbitz, C.~H.} \& \bibinfo{author}{Dalhus, B.}
		\newblock \bibinfo{title}{{{\sc l}-Isoleucine, Redetermination at 120K}}.
		\newblock \emph{\bibinfo{journal}{Acta Crystallographica Section C}}
		\textbf{\bibinfo{volume}{52}}, \bibinfo{pages}{1464--1466}
		(\bibinfo{year}{1996}).
		
		\bibitem{KIMKUO}
		\bibinfo{author}{Pizzarello, S.} \& \bibinfo{author}{Groy, T.~L.}
		\newblock \bibinfo{title}{Molecular asymmetry in extraterrestrial organic
			chemistry: An analytical perspective}.
		\newblock \emph{\bibinfo{journal}{Geochimica et Cosmochimica Acta}}
		\textbf{\bibinfo{volume}{75}}, \bibinfo{pages}{645 -- 656}
		(\bibinfo{year}{2011}).
		
		\bibitem{LEUCIN04}
		\bibinfo{author}{Binns, J.}, \bibinfo{author}{Parsons, S.} \&
		\bibinfo{author}{McIntyre, G.~J.}
		\newblock \bibinfo{title}{{Accurate hydrogen parameters for the amino acid {\sc
					l}-leucine}}.
		\newblock \emph{\bibinfo{journal}{Acta Crystallographica Section B}}
		\textbf{\bibinfo{volume}{72}}, \bibinfo{pages}{885--892}
		(\bibinfo{year}{2016}).
		
		\bibitem{CUFFUG}
		\bibinfo{author}{Williams, P.~A.}, \bibinfo{author}{Hughes, C.~E.} \&
		\bibinfo{author}{Harris, K. D.~M.}
		\newblock \bibinfo{title}{L-lysine: Exploiting powder x-ray diffraction to
			complete the set of crystal structures of the 20 directly encoded
			proteinogenic amino acids}.
		\newblock \emph{\bibinfo{journal}{Angewandte Chemie International Edition}}
		\textbf{\bibinfo{volume}{54}}, \bibinfo{pages}{3973--3977}
		(\bibinfo{year}{2015}).
		
		\bibitem{LYSCLH11}
		\bibinfo{author}{Koetzle, T.~F.}, \bibinfo{author}{Lehmann, M.~S.},
		\bibinfo{author}{Verbist, J.~J.} \& \bibinfo{author}{Hamilton, W.~C.}
		\newblock \bibinfo{title}{{Precision neutron diffraction structure
				determination of protein and nucleic acid components. VII. The crystal and
				molecular structure of the amino acid {\sc l}-lysine monohydrochloride
				dihydrate}}.
		\newblock \emph{\bibinfo{journal}{Acta Crystallographica Section B}}
		\textbf{\bibinfo{volume}{28}}, \bibinfo{pages}{3207--3214}
		(\bibinfo{year}{1972}).
		
		\bibitem{LMETON14}
		\bibinfo{author}{G{\"{o}}rbitz, C.~H.}, \bibinfo{author}{Karen, P.},
		\bibinfo{author}{Du{\v{s}}ek, M.} \&
		\bibinfo{author}{Pet{\v{r}}{\'\i}{\v{c}}ek, V.}
		\newblock \bibinfo{title}{{An exceptional series of phase transitions in
				hydrophobic amino acids with linear side chains}}.
		\newblock \emph{\bibinfo{journal}{IUCrJ}} \textbf{\bibinfo{volume}{3}},
		\bibinfo{pages}{341--353} (\bibinfo{year}{2016}).
		
		\bibitem{VUKQID}
		\bibinfo{author}{Arkhipov, S.~G.}, \bibinfo{author}{Rychkov, D.~A.},
		\bibinfo{author}{Pugachev, A.~M.} \& \bibinfo{author}{Boldyreva, E.~V.}
		\newblock \bibinfo{title}{{New hydro{\-}phobic {\sc l}-amino acid salts:
				maleates of {\sc l}-leucine, {\sc l}-isoleucine and {\sc l}-norvaline}}.
		\newblock \emph{\bibinfo{journal}{Acta Crystallographica Section C}}
		\textbf{\bibinfo{volume}{71}}, \bibinfo{pages}{584--592}
		(\bibinfo{year}{2015}).
		
		\bibitem{QQQAUJ07}
		\bibinfo{author}{Ihlefeldt, F.~S.}, \bibinfo{author}{Pettersen, F.~B.},
		\bibinfo{author}{vonBonin, A.}, \bibinfo{author}{Zawadzka, M.} \&
		\bibinfo{author}{G\"{o}rbitz, C.~H.}
		\newblock \bibinfo{title}{The polymorphs of l-phenylalanine}.
		\newblock \emph{\bibinfo{journal}{Angewandte Chemie International Edition}}
		\textbf{\bibinfo{volume}{53}}, \bibinfo{pages}{13600--13604}
		(\bibinfo{year}{2014}).
		
		\bibitem{PROLIN01}
		\bibinfo{author}{Seijas, L.~E.}, \bibinfo{author}{Delgado, G.~E.},
		\bibinfo{author}{Mora, A.~J.}, \bibinfo{author}{Fitch, A.~N.} \&
		\bibinfo{author}{Brunelli, M.}
		\newblock \bibinfo{title}{On the crystal structures and hydrogen bond patterns
			in proline pseudopolymorphs}.
		\newblock \emph{\bibinfo{journal}{Powder Diffraction}}
		\textbf{\bibinfo{volume}{25}}, \bibinfo{pages}{235–240}
		(\bibinfo{year}{2010}).
		
		\bibitem{LSERIN28}
		\bibinfo{author}{Boldyreva, E.} \emph{et~al.}
		\newblock \bibinfo{title}{Pressure-induced phase transitions in crystalline
			l-serine studied by single-crystal and high-resolution powder x-ray
			diffraction}.
		\newblock \emph{\bibinfo{journal}{Chemical Physics Letters}}
		\textbf{\bibinfo{volume}{429}}, \bibinfo{pages}{474 -- 478}
		(\bibinfo{year}{2006}).
		
		\bibitem{LTHREO04}
		\bibinfo{author}{Woi{\'n}ska, M.}, \bibinfo{author}{Grabowsky, S.},
		\bibinfo{author}{Dominiak, P.~M.}, \bibinfo{author}{Wo{\'z}niak, K.} \&
		\bibinfo{author}{Jayatilaka, D.}
		\newblock \bibinfo{title}{Hydrogen atoms can be located accurately and
			precisely by x-ray crystallography}.
		\newblock \emph{\bibinfo{journal}{Science Advances}}
		\textbf{\bibinfo{volume}{2}} (\bibinfo{year}{2016}).
		
		\bibitem{VIXQOK01}
		\bibinfo{author}{G{\"{o}}rbitz, C.~H.}, \bibinfo{author}{T{\"{o}}rnroos, K.~W.}
		\& \bibinfo{author}{Day, G.~M.}
		\newblock \bibinfo{title}{{Single-crystal investigation of {\sc l}-tryptophan
				with {\it Z}{$^\prime$} = 16}}.
		\newblock \emph{\bibinfo{journal}{Acta Crystallographica Section B}}
		\textbf{\bibinfo{volume}{68}}, \bibinfo{pages}{549--557}
		(\bibinfo{year}{2012}).
		
		\bibitem{LTYROS11}
		\bibinfo{author}{Frey, M.~N.}, \bibinfo{author}{Koetzle, T.~F.},
		\bibinfo{author}{Lehmann, M.~S.} \& \bibinfo{author}{Hamilton, W.~C.}
		\newblock \bibinfo{title}{Precision neutron diffraction structure determination
			of protein and nucleic acid components. x. a comparison between the crystal
			and molecular structures of l\-tyrosine and l\-tyrosine hydrochloride}.
		\newblock \emph{\bibinfo{journal}{The Journal of Chemical Physics}}
		\textbf{\bibinfo{volume}{58}}, \bibinfo{pages}{2547--2556}
		(\bibinfo{year}{1973}).
		
		\bibitem{LVALIN05}
		\bibinfo{author}{Rychkov, D.}, \bibinfo{author}{Arkhipov, S.} \&
		\bibinfo{author}{Boldyreva, E.}
		\newblock \bibinfo{title}{{Structure-forming units of amino acid maleates. Case
				study of {\sc l}-valinium hydrogen maleate}}.
		\newblock \emph{\bibinfo{journal}{Acta Crystallographica Section B}}
		\textbf{\bibinfo{volume}{72}}, \bibinfo{pages}{160--163}
		(\bibinfo{year}{2016}).
		
		\bibitem{VALEHC11}
		\bibinfo{author}{Koetzle, T.~F.}, \bibinfo{author}{Golic, L.},
		\bibinfo{author}{Lehmann, M.~S.}, \bibinfo{author}{Verbist, J.~J.} \&
		\bibinfo{author}{Hamilton, W.~C.}
		\newblock \bibinfo{title}{Precision neutron diffraction structure determination
			of protein and nucleic acid components. xv. crystal and molecular structure
			of the amino acid l\-valine hydrochloride}.
		\newblock \emph{\bibinfo{journal}{The Journal of Chemical Physics}}
		\textbf{\bibinfo{volume}{60}}, \bibinfo{pages}{4690--4696}
		(\bibinfo{year}{1974}).
		
		\bibitem{pdb}
		\bibinfo{author}{Berman, H.~M.} \emph{et~al.}
		\newblock \bibinfo{title}{The protein data bank}.
		\newblock \emph{\bibinfo{journal}{Nucleic Acids Research}}
		\textbf{\bibinfo{volume}{28}}, \bibinfo{pages}{235 -- 242}
		(\bibinfo{year}{2000}).
		
	\end{thebibliography}

	\section*{Acknowledgements}
	MAF’s work was supported by the National Astronomical Observatory of Japan,
	and by a WMU Faculty Research and Creative Activities Award (FRACAA).
	TK was supported partially by Grants-in-Aid for Scientific Research of JSPS (15H03665, 17K05459).
	
	\section*{Author contributions statement}
	M.A.F. provided the quantum chemical calculation results for this project.  R.N.B. and M.A.F. developed the initial
	model for this work.  T.K. provided input into the nuclear physics results and computations.  T.O. provided
	input into the astronomical sites possible in this model.  Y.M. provided expertise in the quantum chemistry codes
	used.  All authors reviewed the manuscript.
	
	\section*{Additional information}
	\begin{itemize}
		\item \textbf{Competing Interests:} The authors declare no competing interests.
		\item \textbf{Correspondence:} Correspondence and requests for materials 
		should be addressed to M.A.F.\\(email: michael.famiano@wmich.edu).
	\end{itemize}
	
	\begin{table}
		
		\centering
		\begin{tabular}{|l|c|c|c|c|c|}
			\hline
			\textbf{Amino Acid}&\textbf{Ligand}&\textbf{Zwitterion}&\textbf{Optimized}&\textbf{$EE_{max}$($\times 10^{-6}$)}&\textbf{Reference}\\
			\hline
			\hline
			Alanine&-3.87&31.79&39.39&2.52&LALNIN59\cite{LALNIN59}\\
			&&&\textbf{51.60}&\textbf{3.35}&XIYSAA\cite{XIYSAA}\\\hline
			Arginine&7.79&-44.11&-160.41&-10.4&TAQBIY\cite{TAQBIY}\\
			&&\textbf{18.57, 47.18}&&\textbf{1.2}&ARGHCL11\cite{ARGHCL11}\\\hline
			Asparagine&21.88&0.05&-36.39&-2.36&VIKKEG\cite{VIKKEG}\\
			&&38.33&37.53&2.38&ASPARM09\cite{ASPARM09}\\\hline
			Aspartic Acid&11.22&11.83&23.66&1.53&LASPRT04\cite{LASPRT04}\\
			&&\underline{15.14}&&\underline{0.98}&NAGLYB10\cite{NAGLYB10}\\\hline
			Cysteine&10.24&8.92&10.60&0.688&LCYSTN36\cite{LCYSTN36}\\\hline
			Glutamic Acid&0.94&28.48&26.54&1.72&LGLUAC03\cite{LGLUAC03}\\
			&&\underline{153.95}&&\underline{10.00}&CAGLCL10\cite{CAGLCL10}\\\hline
			Glutamine&-4.76&16.43&9.22&0.60&GLUTAM02\cite{GLUTAM02}\\
			\hline
			Histidine&-10.55&-44.58&-31.20&-2.02&LHISTD13\cite{LHISTD13}\\
			&&\textbf{20.21}&&\textbf{1.31}&HISTCM12\cite{HISTCM12}\\
			\hline
			Isoleucine&5.67&-5.24, 28.00&17.58&1.14&LISLEU02\cite{LISLEU02}\\\hline
			Isovaline&-0.63&-1.92&-16.67&-1.08&KIMKUO\cite{KIMKUO}
			\\
			~&&&\textbf{119.94}&\textbf{7.68}&\\\hline	
			Leucine&1.95&34.78&30.91&2.01&LEUCIN04\cite{LEUCIN04}\\\hline
			Lysine&0.53&0.09, 42.92&19.78&1.28&CUFFUG\cite{CUFFUG}\\
			&&\textbf{-14.82}&&\textbf{-0.96}&LYSCLH11\cite{LYSCLH11}\\\hline
			Methionine&-1.52&-0.34, 19.09&26.73&1.74&LMETON14\cite{LMETON14}\\\hline
			Norvaline&5.49&26.24&33.22&2.16&USOHUH04\cite{LMETON14}\\
			&&&\textbf{10.50}&\textbf{0.68}&VUKQID\cite{VUKQID}\\\hline
			Phenylalanine&12.10&19.3&21.15&1.37&QQQAUJ07\cite{QQQAUJ07}\\\hline
			Proline&-3.68&17.5&47.25&3.07&PROLIN01\cite{PROLIN01}\\\hline
			Serine&6.84&11.83&13.53&0.88&LSERIN28\cite{LSERIN28}\\\hline
			Threonine&-6.43&12.20&-6.24&-0.41&LTHREO04\cite{LTHREO04}\\\hline
			Tryptophan&18.02&6.51&1.59&0.10&VIXQOK01\cite{VIXQOK01}\\\hline
			Tyrosine&19.72&28.37&-8.90&-0.57&LTYROS11\cite{LTYROS11}\\\hline
			Valine&1.01&4.44, 34.52&19.87&1.29&LVALIN05\cite{LVALIN05}\\
			&&&\textbf{8.47}&\textbf{0.55}&VALEHC11\cite{VALEHC11}\\\hline
			\hline
		\end{tabular}
		\caption{\label{eta_table}Complete table of the molecular geometry parameter $\eta_M$ for amino
			acids not mentioned in the text.
			Also included are geometry references and citations for amino-acids computed in this model.
			Ligand forms were taken from the PDB \cite{pdb}, and zwitterionic and optimized geometries
			are taken from the CSD \cite{csd}.
			Values are in ppm-a.u.  Cations are in 
			\textbf{bold}, and anions are \underline{underlined}. Values separated by commas indicate
			two indicated geometries in the same reference (e.g., two possible
			bond angles in the carboxyl group from the specified geometry).  
			The maximum produced $ee$, $EE_{max}$ is for a meteoroid with v/c=0.015 in a 30 T field at T=10 K and $f$=5$\times$10$^{-9}$.
		}
	\end{table} 
	
	\begin{table}
	\centering
	\begin{tabular}{|l|c|c|c|c|c|c|}
	\hline
	~&\textbf{Ligand}&\textbf{Zwitterion}&\textbf{Optimized}&\textbf{CI1}&\textbf{CM2}&\textbf{CR3}\\
	~&&&\textbf{Cation}&&&\\
	\hline
	\hline
	Isovaline & 0.16 & -0.06 & 2.20 & 1.92(0.16)& 2.45(0.16)&0.05(0.14)\\
	Norvaline & -1.42 & 0.83 & 0.20 & 0.45(0.17) & 0.0(0.14)&0.05(0.14)\\
	Aspartic Acid$^*$& -2.90&0.37&0.60 &0.12(0.17) &2.00(0.15)&-0.10(0.15)\\
	Glutamic Acid$^*$&-0.24 & 0.52&0.73 &3.38(0.17) &2.45(0.17)&-0.10(0.18)\\
	\hline
	\end{tabular}
	\caption{\label{ratios}
		Ratios of the geometry factor for three amino acids (in \%) to that of alanine in this model compared to the equivalent ratios of $ee$s found in the meteoric samples with the most aqueous alteration (CI1),
		the least aqueous alteration (CR3), and an intermediate amount of aqueous alteration (CM2) \cite{glavin10}.
		 Values in parentheses are uncertainties estimated from computed error propagation from the uncertainties provided with the original data.  \\
		  	$^*$Optimized zwitterion geometries used in place of the optimized cation.  
	}
	\end{table}

\end{document}